\documentclass[12pt,preprint]{aastex}

\def\lsim{\lower.5ex\hbox{$\; \buildrel < \over \sim \;$}}
\def\gsim{\lower.5ex\hbox{$\; \buildrel > \over \sim \;$}}

\begin{document}

\title{The Apparent Host Galaxy of PKS 1413+135:  {\it HST}, {\it ASCA}  and
VLBA Observations}

\author{Eric S. Perlman\altaffilmark{1,2}, John T. Stocke\altaffilmark{3}, 
Chris L. Carilli\altaffilmark{4}, Masahiko Sugiho\altaffilmark{5}, Makoto
Tashiro\altaffilmark{6}, Greg Madejski\altaffilmark{7},
Q. Daniel Wang\altaffilmark{8}, John Conway\altaffilmark{9}}

\altaffiltext{1}{Department of Physics, University of Maryland -
  Baltimore County, 1000 Hilltop Circle, Baltimore, MD 21250, USA}
\altaffiltext{2}{Department of Physics and Astronomy, Johns Hopkins
  University, 3400 North Charles Street, Baltimore, MD 21218, USA}
\altaffiltext{3}{Center for Astrophysics and Space Astronomy, University
  of Colorado, Campus Box 389, Boulder, CO  80309, USA}
\altaffiltext{4}{National Radio Astronomy Observatory, P. O. Box 0,
  Socorro, NM  87801, USA}
\altaffiltext{5}{Makishima Laboratory, Department of Physics, University
  of Tokyo,  Hongo 7-3-1, Bunkyo-ku, Tokyo, 113-0033, Japan}
\altaffiltext{6}{Department of Physics, Saitama University, 255 Shimo-Okubo,
  Saitama, 338-8570, Japan}
\altaffiltext{7}{Stanford Linear Accelerator Center, GLAST Group, 2575 Sand 
Hill Road, MS 43A, Menlo Park, CA  94025, USA} 
\altaffiltext{8}{Department of Astronomy, University of Massachusetts, LGRT-B
  619E, 710 North Pleasant Street, Amherst, MA  01003-9305, USA}
\altaffiltext{9}{Onsala Space Observatory, S-43992 Onsala, Sweden} 

\begin{abstract}

PKS 1413+135 ($z=0.24671$) is one of very few radio-loud AGN with an apparent
spiral host galaxy.  Previous authors have attributed its nearly exponential
infrared cutoff to heavy absorption but have been unable to place tight limits
on the absorber or its location in the optical galaxy.  In addition, doubts
remain about the relationship of the AGN to the optical galaxy given the
observed lack of re-emitted radiation.  We present new {\it HST}, {\it ASCA}
and {\it VLBA} observations which throw significant new light on these issues. 
The {\it HST} observations reveal that the active nucleus of PKS 1413+135 has
an extremely red color: $V-H = 6.9$ mag, requiring both a spectral turnover at
a few microns due to synchrotron aging and a giant molecular cloud-sized
absorbing region.  Combining constraints from the {\it HST} and {\it ASCA} data
we derive an intrinsic column $N_H = 4.6^{+2.1}_{-1.6} \times 10^{22}{\rm~
cm^{-2}}$ and covering fraction $f = 0.12^{+0.07}_{-0.05}$.  The spin
temperature of the molecular absorption lines found by previous authors
suggests that the cloud is located in the disk of the optical galaxy, making
our sightline rather unlikely ($P \sim 2 \times 10^{-4}$).  The properties of
this region appear typical of large giant molecular clouds in our own galaxy. 
The HI absorber appears centered 25 milliarcseconds away from the nucleus,
while the X-ray and nearly all of the molecular absorbers must cover the
nucleus, implying a rather complicated geometry and cloud structure, in
particular requiring a molecular core along our line of sight to the nucleus. 
Interestingly, the {\it HST}/NICMOS data require the AGN to be decentered
relative to the optical galaxy by $13 \pm 4$ milliarcseconds.  This could be
interpreted as suggestive of an AGN location far in the background compared to
the optical galaxy, but it can also be explained by obscuration and/or nuclear
structure, which is more consistent with the observed lack of multiple images.

\end{abstract}

\section{Introduction}

Despite two decades of observations in the radio, infrared, optical and X-rays,
the unusual flat-spectrum, radio loud AGN PKS 1413+135 remains a puzzle.  The
source was first classified as a ``red quasar'' by Rieke et al. (1979), and
Bregman et al. (1981) and Beichman et al. (1981) presented the first analysis
of its broadband spectrum, showing an extreme, essentially exponential cutoff
just blueward of the thermal infrared.  Subsequent observations have shown that
its appearance changes radically as one observes in progressively blueward
bands throughout the near-infrared and optical.  In K band, the source is
highly polarized (16 $\pm$ 3\%, Stocke et al. 1992) and shows a featureless
spectrum (Perlman et al. 1996), both properties typical of BL Lacertae (BL Lac)
objects, a classification which was first applied to PKS 1413+135 by Bregman et
al. (1981) and Beichman et al. (1981). Similarly, ground-based imaging in the H
band reveals a dominant nuclear point source surrounded by a faint nebulosity
(Lamer et al. 1999).  But in the optical, the spectrum of the object is red,
with only stellar features and a very weak [OII] line (Stocke et al. 1992). 
Optical images obtained with both ground-based telescopes and {\it {\it HST}}
reveal a clear spiral galaxy but show no evidence of an active nucleus (Stocke
et al. 1992; McHardy et al. 1991, 1994).

A variety of evidence points to heavy absorption as the cause of this spectral
cutoff.  {\it Einstein} observations require an absorbing column of $N_H > 2
\times 10^{22}{\rm ~cm^{-2}}$ (Stocke et al. 1992). Radio observations reveal a
strong, redshifted 21 cm HI absorption line (Carilli et al. 1992), and a rich
variety of molecular species, including OH, $^{12}$CO, $^{13}$CO, HCN, HCO$^+$,
HNC, and CN (Wiklind \& Combes 1995, 1997; Kanekar \& Chengalur 2002).  And an
optical {\it HST} image reveals a prominent dust lane across the disk midplane
(McHardy et al. 1994).  With such a large absorbing column within a powerful
AGN's host galaxy, one might expect to see evidence for a bright, reradiated
thermal continuum or near-IR emission lines (e.g., NGC 1068, Thompson, Lebofsky
\& Rieke 1978; Rotaciuc et al. 1991).  Yet the object's broadband spectrum
shows no evidence of a thermal ``hump''(Bregman et al. 1981, Beichman et al.
1981, Stocke et al. 1992), and near-infrared spectroscopy (Perlman et al. 1996)
found no evidence for emission lines.

The resulting picture of PKS 1413+135 is both puzzling and incomplete. It was
in the light of these mysteries that Stocke et al. (1992) proposed that the AGN
might be background to the optical galaxy, and perhaps also amplified by
gravitational lensing. But confirmatory evidence for this hypothesis has been
difficult to come by. Ground-based images reveal that the nuclear point source
appears centered within the optical galaxy to within $0.05-0.1''$ (Lamer et
al.  1999, Stocke et al. 1992).  And radio VLBI imaging (Perlman et al. 1994,
1996) shows no evidence of double images, instead revealing an arcuate
morphology reminiscent of wide-angle-tail radio sources, but with a total
linear size of only 200 pc.  However, as was pointed out by Perlman et al. 
(1996), these arguments are not entirely decisive, particularly given the large
scale height of the optical galaxy.

Here we report new {\it HST}, {\it ASCA} and {\it VLBA} observations of PKS
1413+135, which shed considerable light on the nature of the host galaxy and
absorbing medium.  In Section 2, we discuss the details of the observations
themselves and our data reduction procedures. In Section 3, we present the new
{\it HST}/NICMOS image and analyze it along with an archival HST/WFPC
observation.   In Section 4, we discuss new X-ray observations of PKS
1413+135.  In Section 5, we discuss new VLBA observations, in the redshifted 21
cm line as well as continuum observations at 2 cm and 0.7 cm (15 GHz and 43
GHz).  Section 6 discusses the overall impact of these results, particularly as
it affects both our picture of the optical galaxy and its relationship to the
AGN.

Throughout this paper, we assume a redshift of $z=0.24671$ for the optical
galaxy associated with PKS 1413$+$135, as derived from the redshifted HI
absorption (Carilli et al. 1992).  We assume $H_0 = 60 {\rm ~km ~s^{-1}
~Mpc^{-1}}$ and $\Omega_{tot}=1$ throughout, which gives a map scale of 1
arcsecond = 4.05 kiloparsecs.

\section{Observations and Data Reduction}

\subsection{{\it HST} Observations}

We observed PKS 1413+135 with {\it HST}/NICMOS on 25 August 1998, using the
F160W filter (corresponding roughly to $H$ band) and the NIC1 camera.  The
resulting image has a scale of approximately 0.0432 arcseconds per pixel.  The
total integration time was 5120s.  To maximize resolution and minimize the
effect of known instrumental problems such as grot (Sosey \& Bergeron 1999),
bad columns and warm pixels, the NICMOS observations were dithered in a square,
four-position pattern with offsets of $1.5''$ between each pointing.

We obtained from the archive WF/PC 1 observations of PKS 1413+135, taken by I. 
McHardy and collaborators on 26 December 1992, using the Planetary Camera mode,
which has a scale of 0.046 arcseconds per pixel.  The observation was taken
with the F555W filter, corresponding roughly to $V$ band.  The total
integration time was 2523 sec, split between three exposures offset from one
another by $\sim 0.5''$.  The WF/PC observations were analyzed extensively by
McHardy et al. (1994), and we do not attempt to repeat their analysis, except
to compare the $V$ and $H$ band images and create an optical/near-IR color
image.

We reduced both {\it HST} datasets in IRAF using the best recommended flat
fields, darks, biases and illumination correction images. Unequal pedestal
effects in the NICMOS data were eliminated with UNPEDESTAL. The dithered NICMOS
images were combined using DRIZZLE.  In the process, we corrected the image for
the slightly rectangular NICMOS pixels and for geometric distortion using the
best available correction files. The dithered WF/PC images were first
registered, then  combined, using CRREJ. We then smoothed the resulting image
using GAUSS, assuming a Gaussian of $\sigma = 3$ pixels.  Flux calibrated
images were obtained from the reduced image using SYNPHOT.  The images were
rotated so that North is along the y axis using header information + IMLINTRAN
in IRAF.The NICMOS image is shown in Figure 1, with the panels showing two
stretches meant to emphasize (respectively) the unresolved AGN and the galactic
structure.  Figure 2 shows a contour map of the nuclear regions of the NICMOS
image. The WF/PC image is shown in Figure 3.

\subsection{X-ray Observations}

The {\it ASCA} observation of PKS 1413+135 was carried out on 24-25 July 1998. 
The SIS detectors were operated in a 1-CCD mode, and all data were converted to
the BRIGHT mode.  The GIS detectors were operated in the standard PH-nominal
mode.  We used standard {\it ASCA} data selection, which includes rejection of
the data during SAA passages, and when the geomagnetic rigidity was lower than
6 GeV $c^{-1}$.  The only exception to the standard screening criteria was that
we accepted only data when the source was at least 20$^{\rm o}$ from the Earth
limb.  This is because we expected that 1413+135 would be a faint X-ray
emitter, and we wanted to minimize the effect of any potential contamination
from the Earth's atmosphere.  We extracted photon data from the resulting X-ray
images from a circle within 3.5$'$ and 3$'$ radii respectively for the GIS and
SIS detectors, selecting the SIS data corresponding to grades 0, 2, 3, and 4. 
For background, we used regions away from the target, avoiding any other
obvious point sources.

For all instruments, the net exposure time was about 36000 seconds after the
above selection criteria were applied. The net count rates were $0.0080 \pm
0.0012$ ct s$^{-1}$ for SIS0, $0.0060 \pm 0.0012$ ct s$^{-1}$ for SIS1, $0.0062
\pm 0.0011$ ct s$^{-1}$ for GIS2, and $0.0118 \pm 0.012$ ct s$^{-1}$ for GIS3. 
The count rates for SIS0, SIS1 and GIS2 match well; however, the count rate for
GIS3 is somewhat higher.  The most likely reason for this discrepancy is that
the location of the source on GIS3 is closer to the optical axis than on GIS2,
SIS1 or SIS0.  As a result, the observing efficiency was higher in GIS3.

For the subsequent spectral fitting, we grouped the data to have at least 20
counts in each PH bin.  As expected, the source was quite faint; the 2-10 keV
flux inferred from the data using the best-fit absorbed power law model as
below was $\sim 9 \times 10^{-13}$ erg cm$^{-2}$ s$^{-1}$.  Needless to say,
with such low count rates, there was no indication of variability during the
{\it ASCA} observations.  For the X-ray spectral fitting, we prepared the
response matrices using the standard {\it ASCA} tools.  This included the
sisrmg tool for generation of the redistribution matrix for the SIS detectors,
and the {\it ASCA}arf tool for preparation of the effective area files.  The
results we report supercede those of Sugiho et al. (1999), which included an
earlier analysis of the same data.

\subsection{VLBI Observations}

\subsubsection{Redshifted HI line Observations}

{\it VLBA} observations of PKS 1413+135 in the redshifted HI line were carried
out on 8 July 1998.  All ten antennas of the {\it VLBA} were used, for 12 hours
observing time.  The observing frequency was 1.137319 GHz, corresponding to the
rest frequency of the HI 21 cm line at $z=0.24671$ (Carilli et al. 1992). 
Spectral line mode was used, with 256 channels of frequency width 15.625 kHz,
corresponding to a velocity width of 4.1 km/s.  The beam of the {\it VLBA} was
21.53 $\times$ 14.55 milliarcseconds in PA $-7.57^\circ$.

The data were correlated at the {\it VLBA} correlator in Socorro, NM.
Fringe-fitting, calibration and mapping were done in AIPS, using a point-source
model for initial fringe fitting.  {\it A priori} amplitude calibration was
done using the gain and system temperature curves for each station, yielding
correlated flux densities.  Maps were made in each channel; a channel 0
continuum map was made and then subtracted from each channel to produce the
final, continuum-subtracted cube.  Hybrid-mapping procedures were started using
clean components from the 18 cm image of Perlman et al.  (1996) as a starting
point for initial phase calibration.  In subsequent iterations of
self-calibration, we allowed first the phase and then both amplitude and phase
to vary.

The cumulative line profile from these {\it VLBA} observations was discussed in
a different context by Carilli et al. (2001).  We refer the reader to that
paper for a discussion of the cumulative line profile. That paper did not give
absorption maps, which is the subject of our discussion in \S 5.

\subsubsection{High-frequency Continuum Observations}

{\it VLBA} observations of PKS 1413+135 at 15 and 43 GHz were carried out on 16
July 1995.  All ten antennas of the {\it VLBA} were used. The data were
obtained in continuum mode, with 128 MHz bandwidth. Both left and right hand
circularly polarized data were obtained.  These observations were part of a
campaign featuring twice-yearly monitoring, designed to measure proper motions
and high-frequency spectral structure in PKS 1413+135; this time block also
included 8 and 22 GHz observations which will be discussed, along with a
detailed discussion of the source structure, spectral morphology and proper
motions in Langston et al. (2002).  The beam of the VLBA at 15 GHz was 0.96
$\times$ 0.49 milliarcsec in PA $-4.43^\circ$, while at 43 GHz it was 0.34
$\times$ 0.19 milliarcsec in PA $0.87^\circ$.

The data were correlated at the {\it VLBA} correlator in Socorro, NM.
Fringe-fitting, calibration and mapping were done in AIPS, using a point-source
model for initial fringe fitting.  {\it A priori} amplitude calibration was
done using the gain and system temperature curves for each station, yielding
correlated flux densities. Hybrid-mapping procedures were started using a point
source model for initial phase calibration.  In subsequent iterations of
self-calibration, we allowed first the phase and then both amplitude and phase
to vary.  Uniform weighting (Robust = -1 in IMAGR) was used to make the final
images.

\section {Results from the {\it HST} Observations}

As can be seen from Figures 1 and 2, the dominant feature in the NICMOS image
is the AGN point source, which is so bright that its first Airy ring is roughly
twice as bright as the galaxy at the same radius. This is quite different from
the situation for the WF/PC image (Figure 3, McHardy et al. 1994), where no
evidence of an unresolved source is seen.  Also marked on Figure 1  is a faint
companion galaxy about $6''$ away, which was noted by Lamer et al. (1999).

\subsection {Isophotal Analysis of the NICMOS Image}

We extracted isophotes for the host galaxy of PKS 1413+135 using the IRAF tasks
ELLIPSE, BMODEL and IMCALC.  This extraction and analysis was done on the
unrotated image, in order to minimize any effects from  rotating the image to
the North-up configuration.  The results of this procedure are shown in Figure
4a-e.  We exclude from our analysis isophotes with semi-major axis $a<0.3''$
because of the extreme brightness of the unresolved AGN source (Figure 1, left
panel). Indeed, the innermost of the isophotes beyond $0.3''$ still shows some
significant effect from the first Airy ring, and there is also evidence for a
second, and even third Airy ring in the isophotes within $0.7''$.

Figure 4a shows the isophotal profile of the optical galaxy.  We fit this
profile to an exponential model of the form $log_{10} (ADU) = b + m*a$, with
best fit values for the parameters $b=2.385$ and $m=-1.072$.  As can be seen,
the fit to this model is quite good at $0.75<a<2''$.  At smaller radii, the
galaxy's profile visibly flattens, such that the model overshoots. This is a
typical characteristic of spiral galaxies with a significant bulge; the
presence of a significant bulge at $a<0.75''$ (i.e., 3 kpc) was in fact noted
by McHardy et al.  (1994) in their analysis of the WF/PC 1 data.  We have shown
the division between the bulge and disk at $a=0.75''$ in Figure 4a by a dashed
vertical line. The large size of the bulge suggests that the optical galaxy is
a fairly early spiral, perhaps an Sa, as suggested by McHardy et al. (1994). 
The disk itself is rather extensive, as at $a>2''$ (8 kpc), the galaxy's
profile flattens yet further, such that the model overshoots.  This might be a
remnant of past interactions with companions such as the one seen on Figure 1.
The bulge can also be seen in Figure 4b, where we plot isophote ellipticity. As
can be seen, $\epsilon$ varies from near zero at $a<0.4''$ to $\sim 0.75$ at
$a>3''$, increasing nearly monotonically with $a$.  As Figure 4c shows, the
dependence of isophotal PA with semi-major axis appears weak.  This is not at
all atypical of bright spiral galaxies seen at large inclination angle
(Courteau 1996).

Figures 4d and 4e plot the location of the centroid of each isophote. In those
plots, the location of the unresolved point source is denoted by a dashed
line.  It is particularly interesting to examine the points within the inner
$0.75''$, i.e., the bulge.  One would expect that the location of the point
source should coincide with the centroid of the bulge isophotes (the disk
isophotes might be affected by patchiness in the disk and/or dust
obscuration).  However, as can be seen, there appears to be a significant
offset.  We take the most conservative estimate of this offset, generated by
the isophotes at $0.3-0.5''$ , which yields a decentering of $0.013'' \pm
0.004''$ or $53 \pm 16$ pc at $z=0.24671$, relative to the isphote centroids
(as can be seen in Figure 4e we find even larger decentering at larger values
of $a$). The X isophotal centers show much less significant (if any) evidence
of an offset.  The offset can just be seen in Figure 2, which shows that the
nucleus appears just northeast of the center of the galaxy isophotes (marked by
a cross).  This is quite surprising, as all previous images of this object have
shown that the AGN appears to be well centered within the optical galaxy (see
Stocke et al. 1992, Lamer et al.  1999).  Those images were, however,
ground-based and could not detect a decentering of $ 0.013''$.  We tried
several different permutations of initial parameters for ISOPHOTE but this
effect remains present.  An off-center AGN could indicate either that the AGN
is background to the optical galaxy, or alternately if the AGN is hosted by the
optical galaxy, that there is significant nuclear structure aligned with the
AGN or radio lobes, perhaps augmented by any effects due to dust, which is not
completely absent at $H$ band ($A_H/A_V = 0.176$ for $R=3.1$, Mathis 1991).  We
discuss both possibilities in \S 6.

\subsection{WF/PC Image and Optical/Near-IR Colors}

As shown by Figure 3, a rather different picture of this system is seen in the
optical than in the infrared. The archival WF/PC image shows no evidence for a
central point; instead, the most prominent feature on the image is a dust lane
more than 4 arcseconds long, extending roughly along the disk plane of the
galaxy, and traversing the nuclear region.  The dust lane appears to be clearly
resolved with a width $\sim 0.25''$ (1 kpc).  The nuclear regions appear
distinctly peanut-shaped, with a central bar extending for over $0.5''$
perpendicular to the dust lane.  Interestingly, the orientation of the nuclear
bar corresponds fairly closely to that of the milliarcsecond radio structure
(Perlman et al. 1996).

By combining the WF/PC and NICMOS images, we constructed a $V-H$ color image.
To do this, we resampled the NICMOS image to $0.046''$/pix, and registered the
two images by assuming that the unresolved point source seen in the NICMOS
image corresponds to the center of the nuclear bar seen on the WF/PC image
(which also corresponds to the middle of the dust lane).  Even though this
assumption is not supported by an image in an intermediate band, it is
consistent with the large absorbing column required by the {\it Einstein} data
(Stocke et al. 1992) as well as the HI and other radio absorption line
observations (Carilli et al. 1992, Wiklind \& Combes 1997, see also \S 5).

The resulting $V-H$ image is shown in Figure 5.  This image has two dominant
features.  First, the disk midplane is considerably redder than other regions
of the galaxy, with typical values of $V-H \sim 2-2.8$ compared to $V-H \sim
1-1.5$.  This is consistent with the presence of dust, as noted above, and as
in McHardy et al. (1994). Interestingly, the dust disk does not appear in the
$V-H$ image as a sharp feature, instead appearing more gradual, perhaps
indicating a patchy distribution of dust in the optical galaxy's disk.  There
is also significant variation in $V-H$ color along the disk midplane, with
generally redder values closer to the nucleus.  This is a property that shared
with the small companion galaxy $6''$ away, which also appears to have a
'disky' morphology and somewhat redder colors in its midplane and nucleus ($V-H
\sim 1.8-2.2$ compared to $V-H \sim 1-1.4$.)

Second, the nucleus is extremely red ($V-H$ = 6.9), four magnitudes redder than
any other feature in the map.  Given the lack of a nuclear point source in the
optical, it is useful to note that since the central 'bar' is the brightest
feature in the WF/PC image any alternate choice for registering the nucleus
would yield an even more extreme color for the AGN.

This speaks to either extreme absorption, an infrared spectral cutoff, or
both.  In fact, the spectral cutoff implied by the $V-H$ color is so extreme
($\alpha > 2.3$ for $S_\nu \propto \nu^{-\alpha}$) that no synchrotron aging
mechanism can account for the observed spectral cutoff by itself, without
resorting to an unphysical exponential cutoff in the particle distribution. 
Thus we are forced to conclude that a significant and probably dominant factor
in the shape of the infrared spectrum of PKS 1413+135 is extinction.  This
conclusion is supported by the large column implied by the X-ray spectrum
(Stocke et al. 1992, \S 4).  However, it is also difficult to account for the
observed nuclear spectrum by reddening alone, given a normal extinction law.
This difficulty was first pointed out by Beichman et al. (1981) without the
X-ray data or a high-resolution near-IR image; given the extreme color we find
for the nucleus it is even more acute now.  The most consistent explanation is
a combination of a rollover due to synchrotron aging combined with extinction,
as first advocated by Stocke et al. (1992).  If, for example, the intrinsic
nuclear spectral index were to steepen by $\Delta \alpha=0.5 $ in the
neighborhood of 3-5 microns ($\Delta \alpha=0.5 $ is predicted for synchrotron
aging with continuous reinjection of electrons, Meisenheimer \& Heavens 1989),
then our data require an excess 3 mag of extinction at the position of the
nucleus.  If one then assumes a normal extinction law (Mathis 1991) and $R_V =
3.1$, the required column is $N_H= 5.7 \times 10^{21} ~{\rm cm^{-2}}$, assuming
a covering factor $f=1$.   A more likely value for the covering factor. 
however, is $\sim 0.1-0.2$, given the value of $N_H$ we derive in \S 4 for the
X-ray absorption (see also Stocke et al. 1992).  This implies 10-30 mag of
extinction along our sightline to the AGN.  These conclusions would be helped
significantly by an {\it HST} image in an intermediate band to confirm the
spectral slope and compare in detail to extinction laws.

\section {ASCA Observations}

PKS 1413+135 has been the target of several X-ray observations.  The source was
quite faint for both {\it Einstein} (Stocke et al. 1992) and {\it ROSAT}
(McHardy et al. 1994), where only photons at energies $>1 $ keV were observed
(only about 15 above background in each case). However, neither {\it Einstein}
nor {\it ROSAT} had significant sensitivity above $\sim 3$ keV, where most of
the observable X-ray emissions for such a heavily absorbed source would be. It
was for this reason that we observed PKS 1413+135 with {\it ASCA}.

For the purposes of this paper, {\it ASCA} is essentially a spectrometer only,
as its angular resolution is $\sim 1 $ arcminute. We show in Figure 6a the
X-ray spectrum of PKS 1413+135 extracted from both the SIS and GIS data. In
Figure 6b we show contour ellipses in the $(N_H,\alpha_x)$ plane at 68 \%, 95\%
and 99\% confidence.  The model fitted in Figure 6 is a power law with two
components of Morrison \& McCammon (1992) absorption.  We fix the Galactic
absorption at the value indicated by the survey data of Stark et al. (1992),
$N_H$(Galactic) $= 2.3 \times 10^{20} {\rm ~cm^{-2}}$, and assume that the
dominant absorption comes from the optical galaxy at $z=0.24671.$ The best fit
parameters for this model are $\alpha_x = 0.66\pm 0.40$ and intrinsic $N_H =
4.6 ^{+2.1}_{-1.6} \times 10^{22} ~ {\rm cm^{-2}}$, and the goodness of fit is
$\chi^2_\nu = 1.068$ (errors are quoted at 90\% confidence).  These parameters
agree well with the parameters fitted to the {\it Einstein} data by Stocke et
al. (1992) as well as those quoted by Sugiho et al. (1999) for an earlier fit
to these data. The X-ray spectral index is typical of those seen in both OVV
quasars and low-frequency peaked BL Lacs (Urry et al. 1996, Sambruna et al.
1999).

The X-ray flux was F (2--10 keV) = $9 \times 10^{-13} {\rm ~erg ~s^{-1}
~cm^{-2}}$, and the derived luminosity was $L(2-10 {\rm ~keV}) = 2.6 \times
10^{44} {\rm ~erg ~s^{-1}}$.  Thus in this observation PKS 1413+135 was fainter
by a factor $\sim 5$ than seen in previous X-ray observations.  It is worth
noting that any 2--10 keV flux figure obtained from the {\it ROSAT} and {\it
Einstein} observations is based almost completely on extrapolations, so the
actual variability could be somewhat different in magnitude.  However, as was
noted by Stocke et al. (1992), Stevens et al. (1994), Perlman et al. (1996) and
Lamer et al. (1999), this source is highly variable in both the radio and
optical, so variations of a factor 5 in the hard X-rays should not be seen as
surprising.

The large column required by all the X-ray observations of PKS 1413+135 is
quite consistent with the observed optical absorbing column (\S 3.2) if the
absorbing material is patchy, with covering fraction $f = 0.12^{+0.07}_{-0.05}$
(the errors are at 90\% confidence and are dominated by the error in the column
fit to the {\it ASCA} data).  This suggests that the absorbing material covers
only $\sim 1/2-1/4$ of a WF/PC pixel, i.e., $\sim 50$ parsecs at $z=0.24671$.
Such a size, patchiness and absorbing column would not be atypical of giant
molecular cloud complexes in our own galaxy (e.g., Orion; Green \& Padman
1993).  Given that the observed CO excitation temperature ($\sim 10$ K; Wiklind
\& Combes 1995, 1997) is more typical of outer-galaxy GMC complexes than a
nuclear cloud (Maloney 1990), it is then useful to point out that our
sight-line to PKS 1413+135 remains highly unusual.  Indeed, if one assumes
projected dimensions of $\sim 1 \times 15$ kpc for the dust lane, the
probability of observing the AGN projected behind a $\sim 50 \times 50$ pc GMC
complex is $\sim 2 \times 10^{-4}$.  Since PKS 1413+135 has unique properties
for a Parkes radio source, this low inferred probability is consistent with the
very small percentage of similar sources found in bright radio surveys.

If indeed a large amount of absorbing material were present in the nuclear
regions, we would expect to see a bright, but narrow, Fe K$\alpha$ line.  As
can be seen (Figure 6a) no line is observed, although due to the low signal to
noise our upper limit on equivalent width is quite modest: 500 eV. Our
nondetection of this line is consistent with the absorbing material being
either well out in the disk of the optical galaxy, or far in the foreground
compared to the AGN (the so called background AGN hypothesis, cf. \S 6).

\section {{\it VLBA} Observations} 

\subsection {HI Absorption Observations}

The existence of a redshifted 21 cm HI line in the spectrum of PKS 1413+135,
discovered by Carilli et al. (1992) was a second important link implicating
significant absorption as the cause of the IR rollover.  In the light of
evidence indicating a patchy absorbing column and a significantly resolved
radio source with a size corresponding to $\lsim 100$ milliarcsec, i.e., $\sim
2$ WF/PC pixels, it is quite useful to look at any spatial structure in the
radio HI absorber.

In Figure 7, we show contour maps of the absorbing material in four of the 256
channels in our {\it VLBA} observations, corresponding to a range of 16 km/s
centered around the frequency of the HI line at $z=0.24671$ (Carilli et al.
1992).  All panels of Figure 7 have the channel 0 image shown in greyscale. 
The nucleus is shown at (0,0) in all four panels.  No other significant
features were found in the image for these or any other channels, although
noise is a significant issue.  For comparison, the single-dish observation of
Carilli et al. (1992) found a FWHM of 18 km s$^{-1}$, consistent with the four
channels found in these observations, but suggesting that some less-obscured
regions may exist in outlying channels, below the noise level. 

Because of the relatively low signal to noise of the absorption maps, we do not
show optical depth maps; however, an examination of those maps indicates that
the optical depth within the absorber is typically $\sim 50-70\%$, i.e.,
optical depth $\tau \approx 1$.  By comparison, Carilli et al. (1992) found a
peak line depth of $0.34 \pm 0.04$. However, those single-dish measurements did
not resolve the source. If, as indicated by Figure 7, a significant amount of
the source flux is not obscured by the HI absorbing screen, multiplying the
observed optical depth by the fraction of flux that is behind the screen,
yields good agreement with the value quoted by Carilli et al. (1992).

As can be seen, the radio HI screen obscures primarily the mini-lobe 15-40
milliarcsec northeast of the nucleus.  There is a possible, slight extension of
the absorber to the southwest (towards the nucleus).  There is also some
marginal velocity structure, with the easternmost part of the screen having a
velocity width somewhat narrower than the region near the mini-lobe's flux
maximum.  We do not observe significant absorption at the nucleus position, to
a $2 \sigma$ optical depth limit of $\sim 0.5$; however as this is not much
lower than the optical depth figures we observe to the eastern mini-lobe this
is only a weak limit. We cannot use the total HI optical depth to improve
significantly on this because of the relative faintness of the nucleus at this
frequency.  Interestingly, the HI absorber covers the regions with the steepest
radio spectral index ($\alpha_r$ ranging from 1.3-2.5) regions in the maps of
Perlman et al. (1996).

\subsection{High-frequency Observations}

The 15 and 43 GHz VLBA images of PKS 1413+135 are shown in Figure 8.  As in
Figure 7, the nucleus is shown at (0,0).  As can be seen, by far the brightest
feature in these images is the nucleus, a stark contrast from the structure
seen in Figure 7 (redshifted 21 cm), where the lobe is a factor 5-6 brighter
than the nucleus.  This is not terribly surprising given the spectral index
maps published by Perlman et al. (1996) which show  a highly inverted spectrum
($\alpha=-1.7$) for the core but very steep spectra for all the extended
structure (ranging from $\alpha=0.7$ to $\alpha > 2$).

The 15 GHz image shows a jet extending west for about 3 milliarcsec, before
taking a bend. There is flux at greater distances which is just barely visible
in the contours but is significant on smoothed image; the position angle of the
jet in this image matches that of the jet at 5 GHz and 8.4 GHz (Perlman et al.
1996). The western jet appears to emerge south of the nucleus's centroid,
indicating a likely second bend closer in to the nucleus which we cannot
resolve. Also visible on the 15 GHz image is a possible counterjet, at a PA
similar to that of the eastern mini-lobe, and about 180$^\circ$ from the
location where the western jet emerges.  At 43 GHz, all that is seen is the
nucleus and a slight extension to the west (supporting the  indications of a
bend at submilliarcsecond scales, seen in the 15 GHz map), which unfortunately
becomes too faint to see about 0.3 milliarcsec from the nucleus.  We see no
evidence of structure outside this fairly simple configuration, and in
particular there is no evidence of a double image of the nucleus down to
resolutions of about 0.2 arcseconds (see \S 6.2). Note that the high frequency
radio emission from PKS 1413+135 is synchrotron in nature and thus physically
unrelated to the high-frequency absorption features superposed along our line
of sight.

\section {Discussion}

The data shown in \S\S 3-5 allow us to place significant constraints on several
of the outstanding mysteries (mentioned in \S 1) regarding the absorbing
material in PKS 1413+135 and its relationship to the AGN.  Based on the NICMOS
image, we confirm that the optical galaxy is indeed a spiral, as found by
previous workers.  The spiral has a fairly large nuclear bulge, about 3 kpc in
size, with a significantly flatter surface brightness profile than the outer
regions of the galaxy.  McHardy et al. (1994) found that the scale height of
the bulge of the optical galaxy was 6.9 kpc, a figure which our data support.

Two issues addressed in the foregoing section particularly bear further
elucidation.

\subsection{Location and Properties of the Absorbing Material}

Our data provide strong evidence that the AGN of PKS 1413+135 is heavily
reddened by absorbing dust and gas along our line of sight.  The optical galaxy
itself has a prominent dust lane along the disk midplane which measures 15 kpc
$\times$ 1 kpc, where the $V-H$ color is about 1 mag redder then elsewhere in
the galaxy.  Such features are common among edge-on spirals.  However, the AGN
has a far more extreme color than anything else in the image ($V-H = 6.9$) ,
implying a spectral cutoff so steep that it requires both an intrinsic break at
a few microns plus 3-4 mag of extinction.  The implied absorbing column is
consistent with the {\it VLBA} HI and {\it ASCA} observations if the covering
factor $f=0.1-0.2$ and the HI spin temperature is a few hundred degrees. 
Combining this with a molecular line spin temperature of $\sim 10 $ K (Wiklind
\& Combes 1995, 1997) we can state that it is most likely that the absorber is
a giant molecular cloud (GMC) complex in the outer reaches of the optical
galaxy's spiral disk (a location in the galaxy's nucleus would predict a
somewhat higher spin temperature and a strong Fe K$\alpha$ line, which is not
observed).  The superposition of an outer-galaxy GMC right along the line of
sight is rather unlikely ($P\approx 2 \times 10^{-4}$, as discussed in \S 4),
but it does appear to be the most consistent hypothesis.

As discussed in \S 5.1, the radio HI absorber appears to be centered about
20-25 milliarcseconds east of the nucleus, along our line of sight to the
eastern mini-lobe.  This mini-lobe dominates the radio flux below about 2 GHz
(Perlman et al. 1996), with a surface brightness some 5-6 times that of the
core at 1139 MHz.  This offset of the HI absorber with respect to the nucleus
(which must be covered by the X-ray absorber) is very interesting.  It
therefore behooves us to inquire as to the relationship of the various
absorption components to one another.  It would not at all be surprising, for
example, to have a GMC complex or star formation region with both warm and cold
components, with the warm component responsible for the X-ray absorption and
the colder, dusty component responsible for the optical and radio absorption.

Absorption lines from a wide variety of molecular species have been observed in
the radio spectrum of PKS 1413+135.  Only one of these lines is at low
frequencies - the recently-discovered OH line found in GMRT observations
(Kanekar \& Chengalur 2002).    The remainder of the lines, discovered by
Wiklind \& Combes (1995, 1997), were all at much higher frequencies (ranging
from about 70-200 GHz in the observed frame; see Table 1 of Wiklind \& Combes
1997).  All the molecular lines are narrow, with the OH absorption width being
14 km $s^{-1}$ while the other lines are $\lsim 10 {\rm ~km ~s^{-1}}$, and only
very slightly offset from the HI line center ($\leq 4 {\rm ~km ~s^{-1}}$
compared to an HI line width of 18 km s$^{-1}$).

As its radio structure spans only 100 millarcseconds (Perlman et al. 1996), PKS
1413+135 was unresolved for all the molecular line observations.  However,
given the information presented here and in Perlman et al. (1996) we have
adequate information to resolve the likely location of the molecular absorbing
material.  At low frequencies, the eastern mini-lobe dominates the radio flux
(Figure 7, \S 5.1).  Therefore, we believe it is more likely that the material
which produces the OH line covers the eastern mini-lobe, although it may also
partially cover the core, a conclusion somewhat at variance with that of
Kanekar \& Chengalur (2002).  The situation is different, however, for the
higher-frequency lines.  Above 20 GHz, the core  accounts for nearly all the
emission from PKS 1413+135 (Figure 8).  Thus the molecular material accounting
for the high-frequency lines observed by Wiklind \& Combes must be projected
within a small distance of the core ($<0.2$ milliarcsec given the compactness
of the 43 GHz structure, Figure 8).

To sum up, then, the center of the observed HI absorber appears to be along the
line of sight to the eastern mini-lobe, 20-25 milliarcseconds from the
nucleus.  However, the X-ray absorbing material and most of the molecular
absorbing material must cover the core, where our data gives a weak $2 \sigma$
upper limit of $\tau_{HI} \approx 0.5$, only $\sim factor 2$ less than that
observed against the lobe. Thus the radio molecular and HI absorbers are {\it
not} necessarily all physically co-located.  Looking at the small velocity
difference between the HI and molecular features, however, it is still most
likely that all the absorbing material is part of the same GMC complex, as much
larger velocity differences would be expected if the HI and molecular clouds
were at different locations within the galaxy's spiral disk.  The projected
location of most of the molecular absorbing material (i.e., covering the VLBI
core) is logical, given the observed optical/near-IR reddening.

With a linear size of at least 40$\times 25$ milliarcseconds (160 $\times $ 100
pc) and an HI velocity width of $18 {\rm ~km ~s^{-1}}$, this GMC complex must
be somewhat more massive than, for example, the Orion region in our own galaxy
(which spans $\sim 100 \times 100$ pc region and has an HI velocity width of a
few km s$^{-1}$), and it likely has a rather patchy and/or filamentary
structure, similar to GMC complexes in our own galaxy.  A patchy structure
would make it rather difficult to estimate a mass or other physical parameters
for the absorber.  For the sake of illustration only we assume a cylindrical
GMC region measuring $50 \times 25 \times 25$ milliarcsec ($200 \times 100
\times 100$ parsecs), with the long dimension superposed near the direction of
the eastern lobe but overlapping the position of the nucleus, and with one of
the molecular cores located virtually along our line of sight to the nucleus
(within 0.2 milliarcsecond or 0.8 parsecs impact parameter - see \S 6.2). Such
a configuration is consistent with the covering fraction constraints from
combining the optical and ASCA data.  To conform with both the HI and molecular
line data it would have to be centered somewhere along the line between the
lobe and nucleus, at perhaps 15 milliarcsec projected distance from the
nucleus.  This works out to a volume of just under $2.1 \times 10^6$ cubic
parsecs, or $6.1 \times 10^{61} {\rm ~cm^3}$.  For a path length of 100 parsecs
($3.09 \times 10^{20}$ cm) and $N_H = 4.6 \times 10^{22}~ {\rm cm^{-2}}$ the
mean density would be $n_H \sim 150~ {\rm cm^{-3}}$, thus yielding a mass of
$\sim 7.6 \times 10^6 M_\odot$.  While this estimate is of course geometry
dependent, it is roughly comparable to the largest GMC complexes in our own
galaxy.  Deeper observations are required to further constrain the
configuration of the radio absorbing material.

\subsection {Relationship of the AGN to the Optical Galaxy and Absorber}

Perhaps the most surprising result of our isophotal analysis is that the {\it
HST} data seem most consistent with an AGN position very slightly offset
($0.013''$, just at the edge of {\it HST}'s astrometrical capabilties in the
near-IR) from the center of the galaxy's isophotes.  This could be interpreted
as the first evidence in favor of the background source hypothesis.  If indeed
the source were decentered, then given the very large scale length of this
galaxy (6.9 kpc) and a smoothly varying projected mass distribution, we might
not expect to see either double images or an arc (see Narayan \& Schneider 1990
although n.b., this source has a decentering two orders of magnitude smaller
than that seen for AO0235+164 and 0537$-$441, the sources modelled in that
paper).

In light of the evidence presented here, however, we cannot assume that the
projected mass distribution of the optical galaxy is smooth.  To the contrary,
a patchy, likely filamentary GMC complex with mass $\sim 7.6 \times 10^6
M_\odot$ and size $\gsim 160 \times 100$ pc must lie along our line of sight to
the AGN.  This geometry is considerably more interesting with respect to the
issue of possible microlensing of a background source and the AGN/absorber
relationship.  The Einstein radius for a $7.6 \times 10^6 M_\odot$ GMC is about
5.7 milliarcsec (for an AGN at $z \sim 1$), compared to a likely projected
location of the cloud center $\sim 25$ milliarcsec from AGN position on the
sky.  Thus the most naive examination of the geometry would conclude that
milli-lensing by the GMC complex is quite unlikely. However, the molecular line
and X-ray absorption data lead us to conclude that there is a molecular core
within 0.2 milliarcsecond of the line of sight to the nucleus.  If the
molecular cloud core is responsible for all the X-ray absorption, and if we
assume a density in this region typical of molecular cores, $\sim 10^3 {\rm
~cm^{-3}}$, the path through the absorber must be $\sim 13$ pc long.  Assuming
this molecular core to be spherical, then, we derive a mass of 32000 $M_\odot$
which must be located within at most 0.2 milliarcsec of our line of sight --
well within its Einstein radius of 0.37 milliarcsec. One would expect such a
lens to produce multiple images with separation $\sim 0.2$ milliarcsec, which
is not seen in Figure 8. Given the available data, the only way for a
background AGN to not produce prominent evidence of microlensing would be to
have the AGN at $z<0.3$.  At such a redshift (only slightly greater than that
of the optical galaxy) the host galaxy of the AGN would easily be visible on
the near-IR image - yet as can be seen from Figure 1 we see no evidence of it. 
Thus the available data now make the background AGN hypothesis even more
unlikely.

There are other, less exotic explanations for such a small decentering.  If
indeed the absorption is patchy and $N_H \approx 4 \times 10^{22} ~ {\rm
cm^{-2}}$ as suggested by our data, some regions of the nucleus might well be
absorbed, even in H band.  This could significantly bias the isophote
centroids, since for a standard extinction law one would still expect 2-5 mag
of extinction at $H$ band for 10-30 mag of extinction at V band.  Furthermore,
if the nuclear ``bar'' is the result of structure aligned and perhaps
associated with the radio source's ``lobes'' (i.e., a miniature alignment
effect, as seen in, for example, 4C31.04, Perlman et al.  2001), such aligned
emission would not be expected to be symmetrical and in fact would be expected
to bias the measurement of isophote centroids.  Either (or both) of these
constitute a much more plausible explanation for the observed decentering given
the available data.

As was previously suggested by Perlman et al. (1996), one can reconcile an
absorber within the optical galaxy with the observed lack of re-emitted AGN
continuum by locating the absorber in the outer disk, as our data seem to
suggest.  One would also have to assume some beaming of the continuum (as in
the models of Snellen et al. 1998) to explain the observed lack of IR emission
lines; however that does not appear to be inconsistent with the known
properties of PKS 1413+135 including high and variable polarization, extreme
variability and extremely core-dominated high-frequency VLBI structure,
although it would require some superluminal motion, which was not seen by
Perlman et al. (1996) based on two-epoch data which, however, did not
adequately resolve the jet's inner regions.  We will return to this latter
issue in Langston et al. (2002, in preparation).  The same model would,
however, require a somewhat underluminous broad line region compared to most
Seyfert galaxies, as observed in other compact symmetric object radio sources
(Perlman et al. 1996 first drew the link between PKS 1413+135 and compact
symmetric objects).  Indeed, the reasonability of this hypothesis is
underscored by the fact that IRAS data in NED (which was not included in
Perlman et al.  1996) show fluxes at 60 $\mu$m and $100 \mu$ m of 220-290 mJy
(two observations with some possible variability between them) and a roughly
flat spectrum between 60-100 $\mu$m.  This is within 50 \% of the prediction
one would make by simply scaling up by a factor of 10 (appropriate to the
observed $N_H$) the correlation of Knapp (1990) between galaxy magnitude and 60
$\mu$m flux.

\subsection{Conclusions}

The most consistent explanation to the properties we observe in PKS 1413+135
appears to be that the absorbing screen consists of a GMC complex within the
outer disk of the optical galaxy.  This geometry is fully consistent with our
new {\it HST, ASCA} and {\it VLBA} observations and is also fully consistent
with previous observations in other wavebands.  The optical galaxy is most
likely the true host of the AGN and radio source given the properties we
observe.  While it is still not possible to fully rule out the background AGN
hypothesis, it now becomes significantly less likely given that the absorption
data require a $3 \times 10^4 M_\odot$ molecular core within 0.2 milliarcsec of
our line of sight.

\vfill\eject

\vfill\eject

\begin{figure}

\plotone{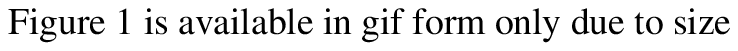}

\caption{ The mosiacked HST NICMOS F160W (H band) image of PKS 1413+135, shown
with two different greyscale stretches.  The scale shown at left, which
emphasizes the nuclear point source, runs from 0 to 4 ADU/s, while the scale
shown at right, which shows the galaxy better, runs from 0 to 1.3 ADU/s.  The
image is shown with a North-up, East-left orientation, and a scale bar is
given.  In addition to the dominant point source, the disk of the optical
galaxy is easily seen, as is the companion galaxy located $6''$ to the East,
which was first noted by Lamer et al. (1999).  See \S 3.1 for discussion.}

\end{figure}
\vfill\eject
\begin{figure}

\plotone{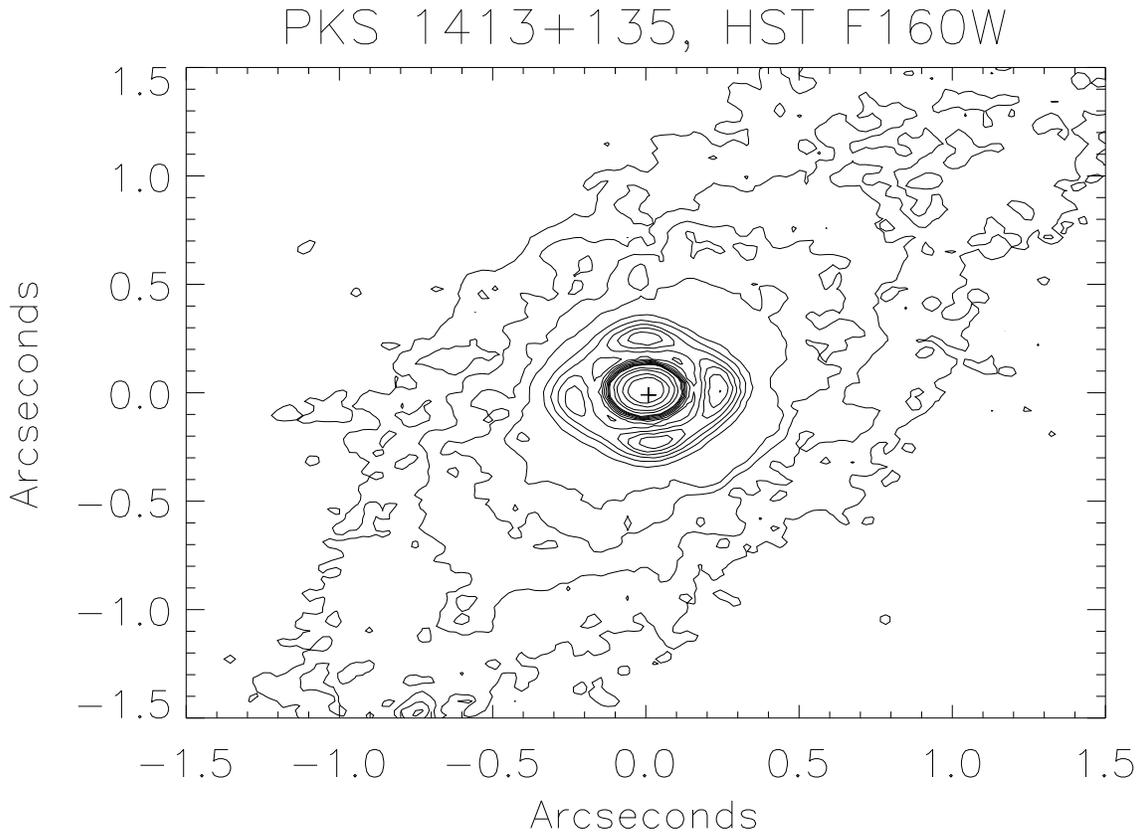}

\caption{A contour plot of the mosaicked HST NICMOS F160W (H band)   image of
PKS 1413+135.  Only the innermost $3 \times 3$ arcseconds are shown. The image
is shown with a North-up, East-left orientation, and the (0,0) point was chosen
to be the location of the centroid of the galaxy isophotes.  Contours are shown
at 120, 200, 300, 400, 800, 1200, 1600, 2000, 2400, 2800, 3200, 3600, 4000,
6000, 8000, 12000, 16000 ADU per pixel.  The brightness of the nuclear point
source is apparent; as can be seen, the first Airy ring is considerably
brighter than the galaxy at a comparable radius.  The nuclear point source
(denoted by a cross) is seen to be offset by about $0.013''$ from the centroid
of the inner galaxy isophotes, in a direction approximately perpendicular to
that of the disk.  See \S 3.1 for discussion.}

\end{figure}
\vfill\eject
\begin{figure}
\epsscale{0.7}

\plotone{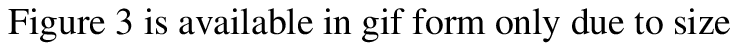}

\caption{The combined HST WF/PC F555W (V band) image of PKS 1413+135.  The
image has been smoothed with a Gaussian of $\sigma = 2 $ pixels to improve the
signal to noise.  Comparing this image with the one shown in Figure 1, the
reader can see the vastly different appearance of PKS 1413+135 in the optical
as compared to the near-IR.  No nuclear point source is apparent in the
optical; instead, the image is dominated by a dust lane which occupies the disk
midplane, and a nuclear bar which extends perpendicular to the dust lane for
about 0.5 arcsec.  See \S 3.2 for discussion.}

\end{figure}
\vfill\eject
\begin{figure}

\epsscale{0.9}
\plotone{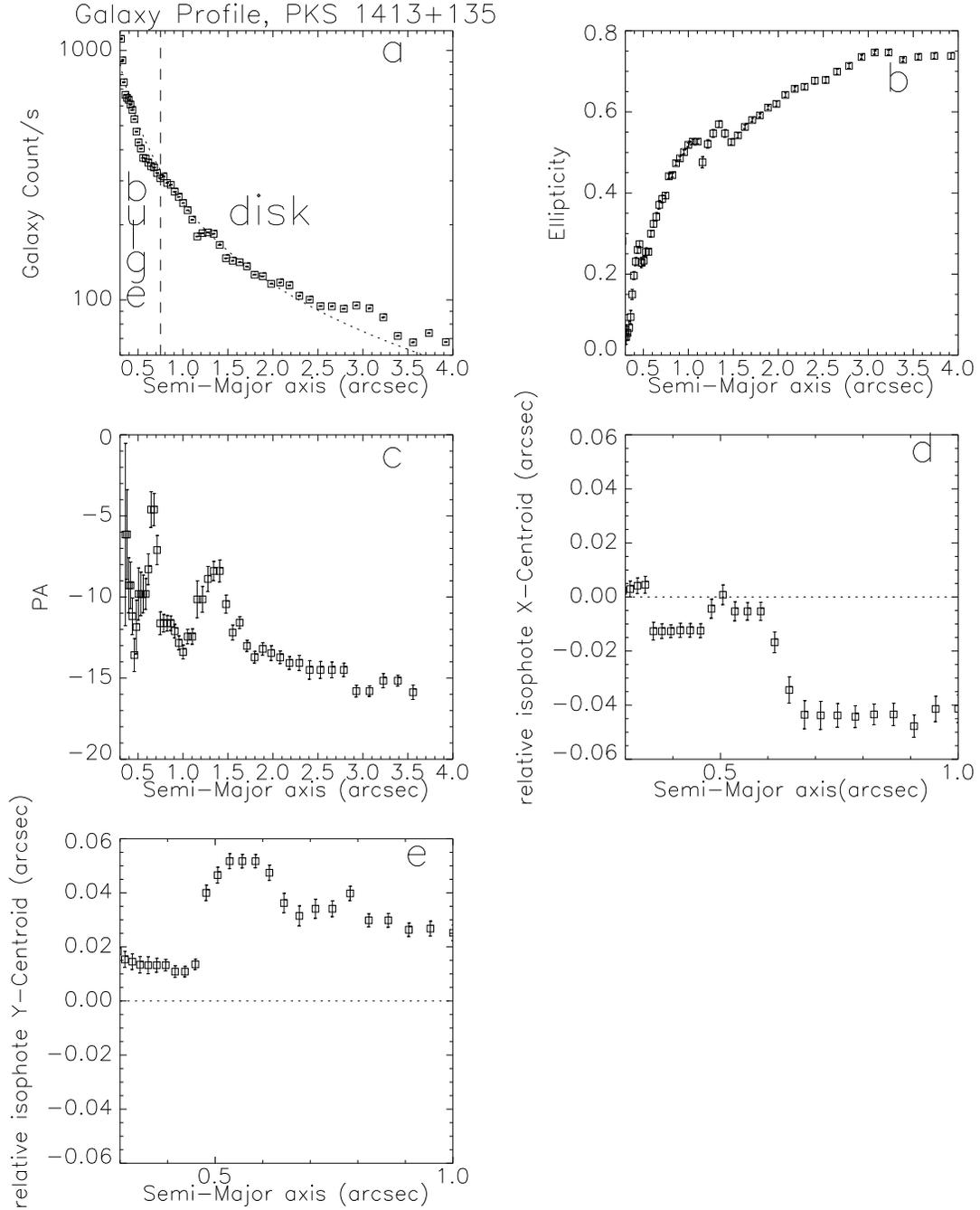}

\caption{ The results of our isophotal analysis.   Five quantities are plotted
versus semi-major axis:  count rate (Fig. 4a), Ellipticity (Fig. 4b), isophotal
position angle (Fig. 4c), isophote X-centroid  (Fig. 4d), and isophote
Y-centroid (Fig 4e).  The dotted line in Fig. 4a denotes the best-fit model,
while in Fig. 4d and 4e it denotes the position of the AGN. We only show
isophotes at $>0.3''$, due to the brightness of the AGN.   This  analysis
was done on the unrotated image; thus $PA=-130^\circ$ here translates to
$PA=0^\circ$ on Figures 1, 2, 3 and 5.  See \S 3.2 for discussion.}

\end{figure}
\vfill\eject
\begin{figure}
\epsscale{0.7}

\plotone{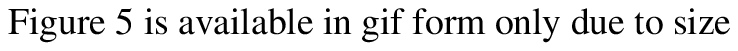}

\caption{  A $V-H$ image of PKS 1413+135, made from the two HST images.  To
make this image, we assumed that the AGN (seen in the near-IR image, Figure 1)
is located at the center of the nuclear bar seen in the optical image (Figure
3).  Darker colors refer to redder regions, and the scale runs from $V-H$ = 1
mag (white) to $V-H$=4 mag (black).  The nucleus, which appears as a saturated
point source in this rendition, is far redder than the limit of the scale
shown, at $V-H = 6.9$ mag. As can be seen, the disk midplane is about 1
magnitude redder than outlying regions of the galaxy, with redder colors seen
closer to the nucleus.  See \S 3.2 for discussion.}

\end{figure}
\vfill\eject
\begin{figure}

\plotone{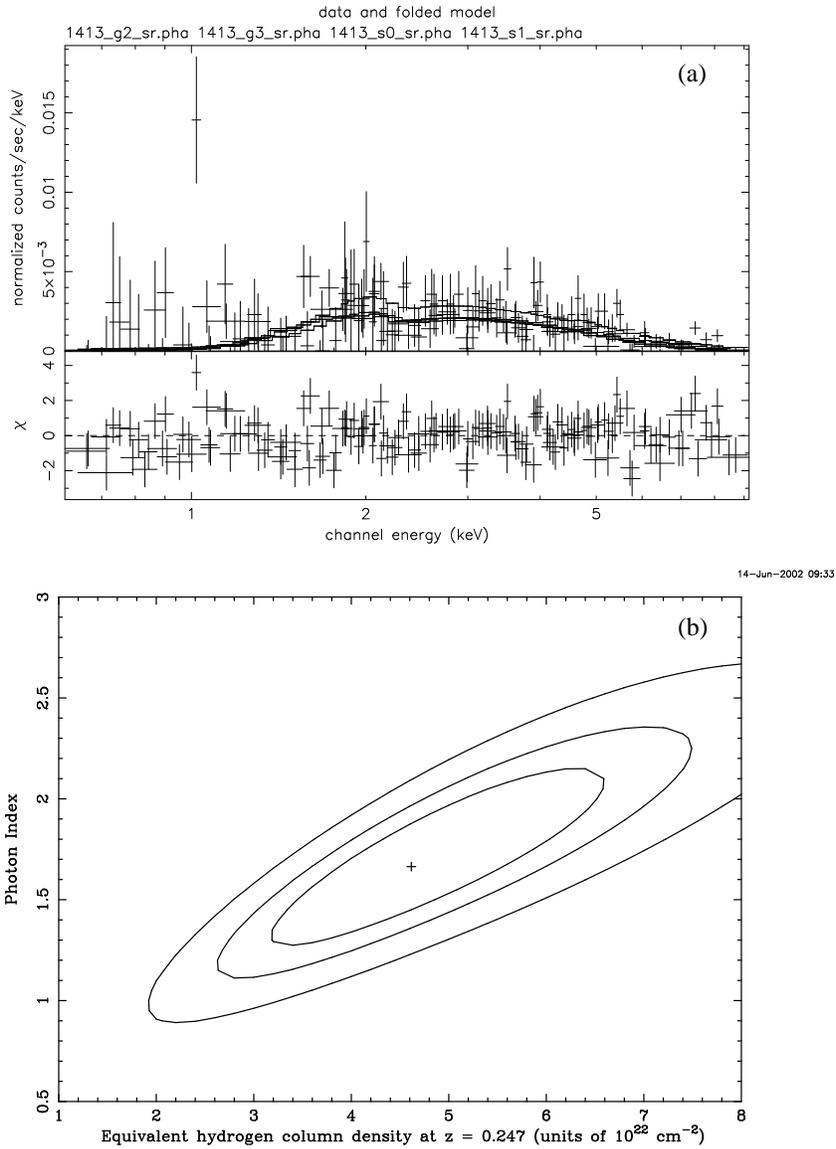}

\caption{ Results of the {\it ASCA} X-ray spectral reduction.  At top, we show
the X-ray spectrum in ct/s/keV plotted versus channel energy in keV, along with
the residual deviations in $\Delta \chi^2$.  Data from all four instruments are
shown individually and fitted simultaneously. The best-fit model is described
in \S 4.  At bottom, we show contours of the best-fit parameters in the
$\Gamma, N_H$ plane.  Contours are shown at the 68\%, 95\% and 99\% confidence
levels.  See \S 4 for discussion.}

\end{figure}
\vfill\eject
\begin{figure}
\epsscale{0.7}
\plotone{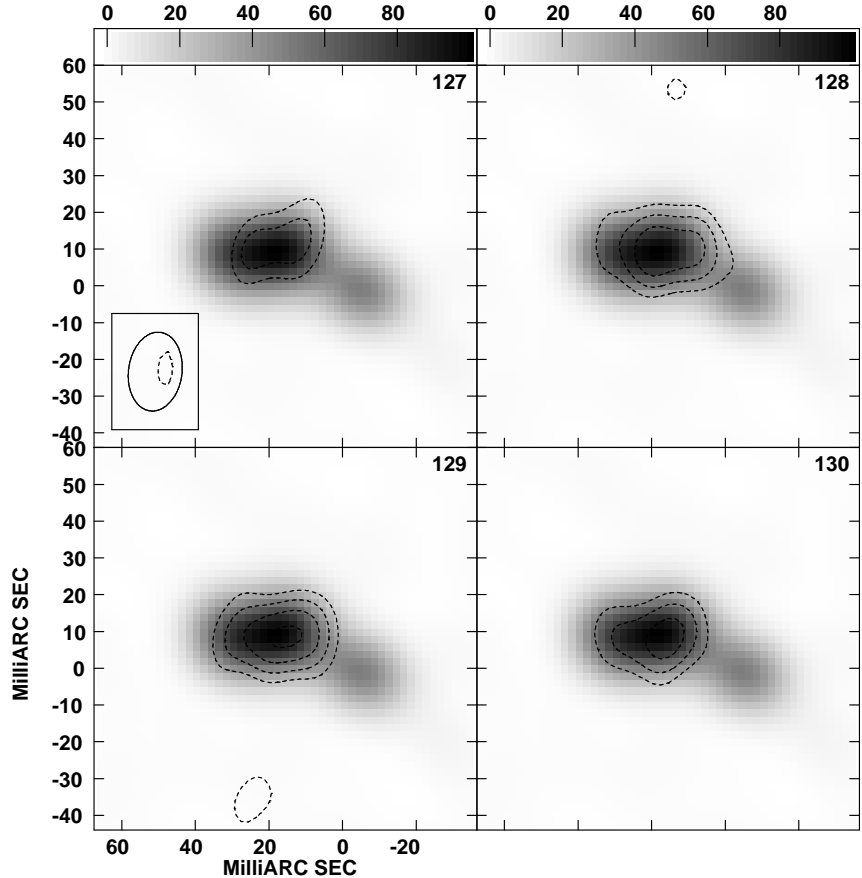}

\caption{ Results of the {\it VLBA} redshifted HI line observations.   Four
channels are shown, each of which is 4.1 km/s wide.  These four channels
correspond to the middle four channels observed in observations centered around
the line frequency found by Carilli et al. (1992). The greyscale image is the
channel 0 image in total flux, while the contours represent deficits due to
absorption. As can be seen, the absorbing material appears to be centered
around the eastern mini-lobe and does not appear to extend as far as the
nucleus (at 0,0).  The peak optical depth against the optical mini-lobe is
approximately 0.7, although the signal-to-noise is fairly low.  Formally, we
can only set a $2 \sigma$ upper limit to the optical depth at the position of
the nucleus of 0.5.  Furthermore, since the flux of the nucleus is a factor six
lower than the eastern mini-lobe at this frequency, we cannot derive
significantly more stringent constraints by assuming a total optical depth
identical to those given in Perlman et al. (1996) and Carilli et al. (2001). 
See \S\S 5, 6 for discussion.}

\end{figure}
\vfill\eject
\begin{figure}

\epsscale{0.65}
\plotone{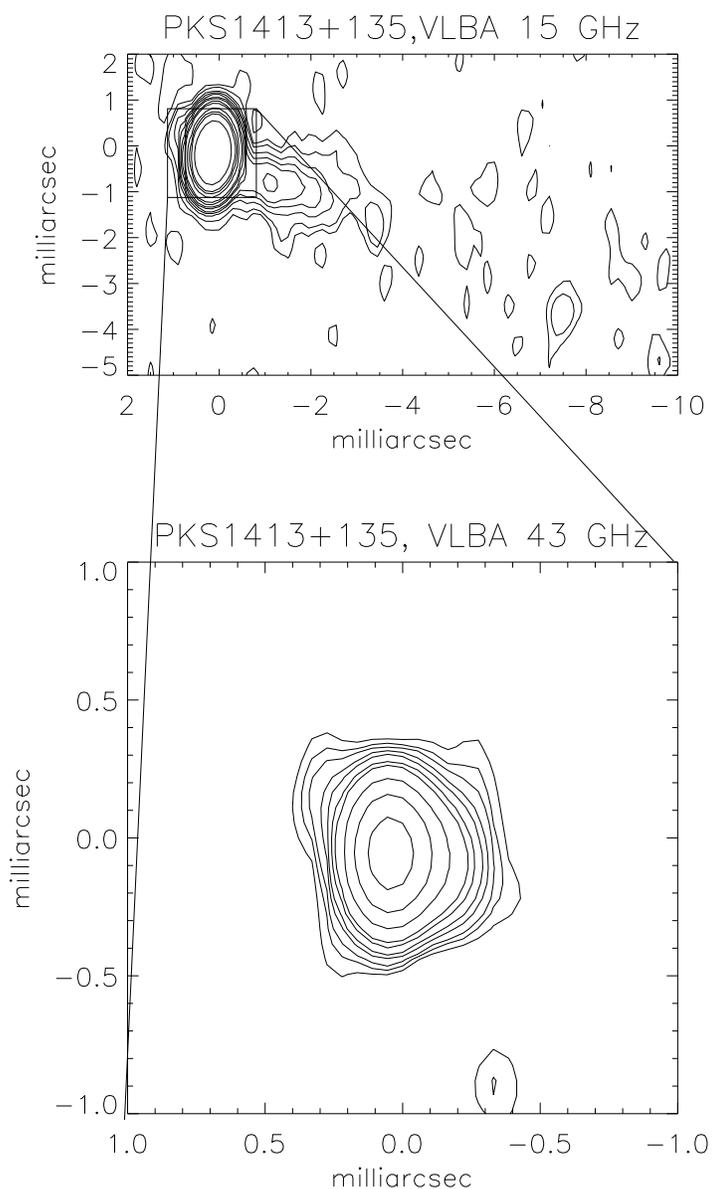}

\caption{Images of PKS 1413+135 at 15 and 43 GHz, obtained with the VLBA.  The
15 GHz map (top, beam = 0.96 $\times$ 0.49 milliarcsec in PA -4.43$^\circ$)
shows a weak jet, with a slight bend at around 3 milliarcsec from the nucleus
(at 0,0).  The very faint structure seen beyond 3 milliarcsec is at the same PA
as the 8 GHz jet seen in Perlman et al. (1996).  There is possible evidence for
a counter-jet as seen on larger scales.  Contours on the 15 GHz image are shown
at (-1 , 1, 2, 4, 8, 12, 16, 24, 32, 64, 96, 128, 256) millijansky per beam. At
43 GHz (bottom, beam = 0.34 $\times$ 0.19 milliarcsec in PA 0.87$^\circ$), the
source is barely extended in the direction of the 15 GHz jet.  Contours on the
43 GHz image are shown at (-8 , 8, 12, 16, 24, 32, 48, 64, 128, 256, 512, 1024)
millijansky per beam. No evidence for double images is seen down to a
resolution of 0.2 milliarcsec.  See \S\S 5, 6 for discussion.}

\end{figure}

\end{document}